\documentstyle[12pt]{article}
\title{
\vspace{-8mm}
\rightline{\small HUB--EP--98/34}
\vspace{-2mm}
\bf String Representation of\\ 
Field Correlators in the\\ 
Dual Abelian Higgs Model}
\author{Dmitri Antonov \thanks{E-mail address: 
antonov@pha2.physik.hu-berlin.de}{\,}
\thanks{On leave of absence from the Institute of Theoretical and 
Experimental Physics, B. Cheremushkinskaya 25, 117 218, Moscow, 
Russia.}~ 
and 
Dietmar Ebert \thanks{E-mail address: debert@qft2.physik.hu-berlin.de}
\\
{\it Institut f\"ur Physik, Humboldt-Universit\"at zu Berlin,}\\
{\it Invalidenstrasse 110, D-10115, Berlin, Germany}}
\date{}
\begin{document}
\maketitle
\vspace{1mm}
\centerline{\bf {Abstract}}
\vspace{3mm}
By making use of the path integral duality transformation, we derive 
the string 
representation for the partition function of an extended Dual Abelian 
Higgs Model containing gauge fields of external currents of 
electrically charged 
particles. By the same method, we obtain the corresponding representations 
for the 
generating functionals of gauge 
field and monopole current correlators. In the case of bilocal correlators,  
the obtained results are found to be  
in agreement 
with the dual Meissner scenario of confinement and with the Stochastic 
Model of the QCD vacuum.

\newpage
{\large \bf 1. Introduction}
\vspace{3mm}

The explanation and description of 
confinement in gauge theories is 
known to be one of the most fundamental problems of modern Quantum 
Field Theory (see e.g. [1,2]). 
In general, by confinement one  
means the phenomenon of absence in the spectrum   
of a certain field theory of the physical $\left|{\rm in}\right>$ and  
$\left|{\rm out}\right>$ states of  
some particles, whose fields are 
however present in the fundamental Lagrangian. 
A natural conjecture here is that due to a linearly rising 
confining interaction, (anti)quarks cannot exist as free particles, 
but form colourless bound states of hadrons.  
The most natural quantity for the description 
of this phenomenon is the so-called Wilson loop 
average. 
In the case of Quantum Chromodynamics (QCD), 
this object has the following 
form  

$$
\left<W(C)\right>=\frac{1}{N_c}\left<{\rm tr}{\,}{\rm P}\exp\left(
ig_{\rm QCD}\oint 
\limits_C^{}A_\mu dx_\mu\right)\right>, \eqno (1)$$
which is nothing else, 
but an averaged amplitude of the process of creation, propagation, 
and annihilation of a quark-antiquark pair. Here,  
$A_\mu$ stands for the matrix-valued vector-potential of the gluonic 
field, $C$ is a closed 
contour, along which the quark-antiquark pair propagates, $P$ stands 
for the path-ordering prescription, and the 
average on both sides is performed with the QCD 
action. The interaction potential of a pair of static quark and antiquark 
can be 
determined from 
the Wilson loop average, defined on the rectangular contour $C$,  
according to the Feynman-Kac formula,

$$V=-\lim\limits_{T\to\infty}^{}\frac{1}{T}\left<W(C)\right>,$$
where $T$ stands for the size of $C$ in the temporal direction. 
In particular, for large contours $C$, the exponential dependence 
of the Wilson 
loop average on the area of the minimal surface, $\Sigma_{\rm min.}$, 
encircled by $C$,

$$
\left<W(C)\right>\to {\rm e}^{-\sigma\cdot \left|\Sigma_{\rm min.}\right|}
$$
(the so-called area law behaviour of the Wilson loop), yields then  
the linearly rising confining potential 

$$
V_{\rm conf.}(R)=\sigma R. \eqno (2)$$
Here $\sigma$ stands for the so-called string tension, and $R$ is the 
relative distance between a quark and an antiquark.  
This exponential asymptotic dependence of the Wilson loop average on 
$\left|\Sigma_{\rm min.}\right|$ is just the essence of the 
Wilson criterion of confinement [3].

The linearly rising potential (2) admits a simple physical interpretation 
in terms of a string picture: 
the gluonic field between 
a quark and an antiquark is compressed to a tube or a string 
(the so-called QCD string), whose 
energy is proportional to its length, and $\sigma$ is the energy of 
such a string per unit length. This string plays the central 
role in the Wilson's picture of confinement, since with 
increasing distance $R$ it stretches and 
prevents quark and antiquark from moving freely to macroscopic distances. 
In order to get an idea of numbers, notice that according 
to the lattice data [4] 
the distance $R$, 
at which the Wilson criterion of confinement becomes valid, is 
of the order of $1.0{\,} {\rm fm}$, and the 
string tension is of the order of $0.2{\,} {\rm GeV}^2$ (see 
e.g. [2]). 

As it is well known, all observable 
dimensional quantities in QCD such as hadron masses, 
and in particular the string tension, 
are proportional to the corresponding power of the QCD scale parameter  
$\Lambda_{\rm QCD}$\footnote{This means that the dimensionless 
ratios of these quantities, e.g. the ratio of $\sqrt{\sigma}$ to 
hadron masses, are universal (i.e. $g_{\rm QCD}$-independent) numbers. 
The aim of all the nonperturbative phenomenological 
approaches to QCD (e.g. the 
so-called Stochastic Vacuum Model, which will be described below) 
is to calculate these numbers, but not 
$\Lambda_{\rm QCD}$ itself.}. In particular, one has 

$$
\sigma\propto\Lambda_{\rm QCD}^2=\mu^2\exp\left[
-\frac{16\pi^2}{\left(\frac{11}{3}N_c-\frac23N_f\right)g_{\rm QCD}^2
(\mu)}\right], \eqno (3)$$
where $g_{\rm QCD}(\mu)$ is the QCD coupling constant measured 
at a scale $\mu$. Eq. (3)  
means that all the coefficients in the expansion of the 
string tension in powers of $g_{\rm QCD}^2(\mu)$ vanish. 
This conclusion tells us 
that the QCD string has a pure 
nonperturbative origin, as well 
as the phenomenon of confinement, which leads to the process of 
formation of such strings in the QCD vacuum, itself.

The problem of studying the 
properties of the QCD string 
is closely related to the 
problem of finding a string representation of gauge theories 
possessing a confining phase.  
In this respect, all methods of derivation of the string 
effective action from the action of gauge theories are of great 
importance. Some progress in the solution of this 
problem has recently been achieved {\it for the case of QCD} in Refs. 
[5-7] using the so-called Stochastic 
Vacuum Model (SVM) of the QCD vacuum [2,8]. 

The idea which 
lies behind this approach is that important quantities like Green 
functions 
can be expressed in terms of the Wilson loop [3,8]. 
The main strategy of SVM is then not to evaluate the 
Wilson loop itself, but to express 
it via gauge-invariant irreducible correlators (the so-called 
cumulants) of the gauge field strength tensors. Such correlators 
have been measured in the lattice experiments both at large and 
small distances [9], which enables one to use them  
in practical calculations of various physical quantities. 
In order to express the Wilson loop average (1) 
via cumulants, one  
makes use of the non-Abelian Stokes theorem and cumulant 
expansion (see [2] for a detailed discussion and Refs. therein), 
which yields

$$
\left<W(C)\right>=
W\left(\Sigma_{\rm min.}\right)\simeq$$

$$\simeq\frac{1}{N_c}{\rm tr}{\,}\exp
\left(-\frac{g_{\rm QCD}^2}{2}\int
\limits_{\Sigma_{\rm min.}}^{}d\sigma_{\mu\nu}(x)
\int\limits_{\Sigma_{\rm min.}}^{}d\sigma_{\lambda\rho}(x')
\left<\left<F_{\mu\nu}(x)\Phi(x,x')
F_{\lambda\rho}(x')\Phi(x',x)\right>\right>\right), 
\eqno (4)$$
where $\Phi(x,y)$ stands for the 
parallel transporter factor along the straight line. 
In the derivation of Eq. (4), 
one has used the so-called bilocal or Gaussian approximation, 
according to which all higher cumulants 
can be disregarded with a good accuracy due to their smallness 
w.r.t. the bilocal cumulant. This approximation is indeed confirmed by the 
present lattice data [9].

The bilocal cumulant standing in the exponent on the R.H.S. of 
Eq. (4) can be parametrized by two renormalization 
group-invariant coefficient functions $D^{\rm QCD}$ and 
$D_1^{\rm QCD}$ as follows [8]

$$
\frac{g_{\rm QCD}^2}{2}\left<\left<F_{\mu\nu}(x)\Phi(x,x')
F_{\lambda\rho}(x')\Phi(x',x)\right>\right>=\frac{\hat 1_{N_c\times 
N_c}}{N_c}\Biggl\{(\delta_{\mu\lambda}\delta_{\nu\rho}-\delta_{\mu\rho}
\delta_{\nu\lambda})D^{\rm QCD}\left((x-x')^2\right)+$$

$$
+\frac12\left[\frac{\partial}{\partial x_\mu}((x-x')_\lambda
\delta_{\nu\rho}-(x-x')_\rho\delta_{\nu\lambda})+\frac{\partial}
{\partial x_\nu}((x-x')_\rho\delta_{\mu\lambda}-(x-x')_\lambda
\delta_{\mu\rho})\right]D_1^{\rm QCD} 
\left((x-x')^2\right)\Biggr\}. \eqno (5)$$

Notice, that by construction 
the term containing the function $D_1^{\rm QCD}$ 
on the R.H.S. of Eq. (5) 
yields a perimeter type 
contribution to the Wilson loop average (4). In particular, the 
leading perturbative 
contribution of the one-gluon-exchange diagram to the Wilson 
loop average is contained in the function $D_1^{\rm QCD}$.

The nonperturbative parts of the functions $D^{\rm QCD}$ and 
$D_1^{\rm QCD}$ have been calculated on the lattice 
in Ref. [9], where it has been 
shown that at large distances $\left|x-x'\right|$,
$D_1^{\rm QCD}\left((x-x')^2\right)\ll D^{\rm QCD}
\left((x-x')^2\right)$,  
and both functions decrease as 
${\rm e}^{-|x-x'|/T_g}$, 
where $T_g$ is the so-called correlation length 
of the vacuum, $T_g\simeq 0.22{\,}{\rm fm}$ in the $SU(3)$-case. 

The idea realized in Ref. [5] was to consider the quantity 
$S_{\rm eff.}=-\ln W\left(\Sigma_{\rm min.}\right)$ 
defined by Eqs. (4) and (5)  
as a starting point for the derivation of the QCD string effective 
action. Due to the smallness of the ratio 
$(T_g/R)^2\simeq 0.04$, it was then possible to perform 
an expansion of the nonlocal action 
$S_{\rm eff.}$ in powers of $(T_g/R)^2$, which yielded the first 
few local terms known in the standard string theory, whose coupling 
constants were completely determined through the bilocal correlator. 
In Ref. [6], perturbative corrections to the obtained string 
effective action arising due to perturbative gluons propagating 
in the nonperturbative QCD vacuum have been calculated, and in 
Ref. [7] a possible way of the solution of the problem 
of crumpling of the string world-sheet has been proposed. 

However, it still remains unclear how one can get a mechanism 
of integration over string world-sheets in QCD, since up to now 
the construction of the string effective action has been performed 
on the surface of the minimal area (cf. Eq. (4)). Elaboration 
out of such a mechanism would eventually completely solve the 
task of finding a string representation of QCD.

The present paper is devoted to be a first step in this direction.   
To this end, we shall simplify the problem under study by considering 
the related Dual Abelian Higgs Model (DAHM). 
This model approach is based on the commonly accepted observation 
that on the phenomenological level,  
quark confinement in QCD can be explained in 
terms of the dual Meissner effect [10,11]. According to the 
't Hooft-Mandelstam picture of confinement, the 
properties of the 
QCD string should be similar to the ones of the electric vortex, which 
emerges between two electrically charged particles immersed 
into the superconducting medium filled with a condensate of 
Cooper pairs of magnetic monopoles. In the case of the usual 
Abelian Higgs Model (AHM), which is a relativistic version of the 
Ginzburg-Landau theory of superconductivity, 
such vortices are 
referred to as Abrikosov-Nielsen-Olesen strings
[12]. 

Thus, it looks 
reasonable to consider DAHM 
as a natural laboratory for probing 
various approaches to the problem of the string representation of QCD 
(for related investigations of AHM see [11,13-18]). 
In this respect, in Ref. [14] the so-called path integral duality 
transformation proposed in Ref. [13] has been applied to the lattice 
version of AHM in the London limit 
in order to reformulate the partition function of this theory 
in terms of the string world-sheet coordinates. 
Then the same reformulation has been performed in the continuum 
limit in Ref. [17].

In this paper, we would like to demonstrate the usefulness 
of the path integral duality transformation for 
the derivation of the string representation for the generating 
functionals 
of gauge field strength tensor and 
monopole 
current correlators in DAHM. In particular, we find it useful 
to study a suitably extended version of this model, which includes 
gauge fields generated by external electrically charged particles 
called ``quarks''.

The outline of the paper is as follows. In Section 2, we shall derive 
the string representation for the partition function of such an 
extended version of DAHM in the London limit.  
The resulting  
action will consist of two parts, the first of which will describe the 
short-range Yukawa interaction between a quark and an antiquark, whereas 
the second one, corresponding to the long-range 
confining potential,  
will contain the integral over all the surfaces encircled by 
the contour of the quark-antiquark pair.

In Section 3, we generalize the result of Section 2 by introducing 
external sources of the field strength tensors in order to derive the 
string representation for the generating functional of gauge field 
correlators in the London limit of extended DAHM. 
In particular, we shall calculate the 
bilocal correlator and demonstrate that its long- and 
short distance asymptotic behaviours are 
in agreement with those found in QCD.

In Section 4, 
we derive the string representation for the generating 
functional of monopole current correlators 
and calculate the correlator of two such currents.

Finally, some technical details concerning the path integral duality 
transformation and the integration over the related Kalb-Ramond field are 
outlined in two Appendices.

\vspace{6mm}
{\large\bf 2. String Representation for the Partition Function of 
the Extended DAHM in the London Limit}
\vspace{3mm}

We shall start with the following expression for the partition 
function of the extended DAHM

$${\cal Z}=
\int \left|\Phi\right| {\cal D}\left|
\Phi\right| {\cal D}B_\mu {\cal D}\theta
\exp\Biggl\{-\int d^4x\Biggl[\frac14 \left(F_{\mu\nu}-
F_{\mu\nu}^E\right)^2+\frac12\left|D_\mu\Phi
\right|^2+\lambda\left(\left|\Phi\right|^2-\eta^2\right)^2\Biggr]
\Biggr\}, \eqno (6)$$
where $\Phi(x)=\left|\Phi(x)\right| {\rm e}^{i\theta(x)}$ is an  
effective Higgs field of ``Cooper pairs'' of magnetic monopoles, 
$B_\mu$ and $F_{\mu\nu}=
\partial_\mu B_\nu-\partial_\nu B_\mu$ are the dual gauge field and 
its field strength tensor,   
$D_\mu=\partial_\mu-2igB_\mu$ is the covariant derivative with 
$g$ standing for the magnetic coupling constant. 
Notice that in Eq. (6), $F_{\mu\nu}^E$ stands for the field 
strength tensor generated by external ``quarks'', defined according to the 
equation 

$$\partial_\nu\tilde F_{\mu\nu}^E\equiv
\frac12\varepsilon_{\mu\nu\lambda\rho}\partial_\nu F_{\lambda\rho}^E=
4\pi j_\mu^E \eqno (7)$$
with

$$j_\mu^E(x)\equiv e\int\limits_0^1 d\tau\frac{dx_\mu(\tau)}{d\tau}
\delta(x-x(\tau))$$
standing for the conserved electric current of a quark,  
which moves along 
the closed contour $C$, parametrized by the function $x_\mu(\tau),{\,}
0\le\tau\le 1,{\,}x_\mu(0)=x_\mu(1)$. The electric coupling 
constant $e$ is related to the magnetic one via Dirac's quantization 
condition 
$eg=\frac{n}{2}$, where $n$ is an integer \footnote{Here, we have adopted 
the notations of Ref. [11].}. 
In what follows, we shall 
for concreteness restrict ourselves to the case of monopoles 
possessing the minimal charge, i.e. put $n=1$.

The solution of Eq. (7) reads 

$$F_{\mu\nu}^E=4\pi e\tilde\Sigma_{\mu\nu},$$
where $\Sigma_{\mu\nu}(x)\equiv\int\limits_{\Sigma}^{}
d\sigma_{\mu\nu}(x(\xi))\delta(x-x(\xi))$ is the so-called vorticity 
tensor current [16] defined on the string world-sheet $\Sigma$. 
This world-sheet is parametrized by the four-vector 
$x_\mu(\xi)$, where  
$\xi=\left(\xi^1, \xi^2\right)$ 
is a two-dimensional coordinate. Due to the Stokes theorem, 
the vorticity tensor current is related to the quark current according 
to the equation 

$$e\partial_\nu\Sigma_{\mu\nu}=j_\mu^E.$$
In particular, 
this equation means, that in the case, when there are no external 
quarks, the vorticity tensor current is conserved, i.e. due to 
the conservation of electric flux all the strings in this case are 
closed. Notice, that 
when external quarks are introduced into the system, 
some amount of closed strings might survive. 
From now on, we shall 
restrict ourselves to the sector of the theory with open strings 
ending at quarks and antiquarks only.     

In the London limit, $\lambda\to\infty$, the radial part of the 
monopole field becomes fixed to 
its v.e.v., $\left|\Phi\right|\to\eta$, and 
the partition function (6) 
takes the form 

$${\cal Z}=\int {\cal D}B_\mu{\cal D}\theta^{{\rm sing.}}{\cal D}
\theta^{{\rm reg.}}\exp\left\{-\int d^4x\left[\frac14
\left(F_{\mu\nu}-F_{\mu\nu}^E\right)^2+\frac{\eta^2}{2}
\left(\partial_\mu\theta-2gB_\mu\right)^2\right]\right\}, \eqno (8)$$
where from now on constant normalization factors will be omitted. 
In Eq. (8), we have performed a 
decomposition of the phase of the magnetic Higgs field $\theta=
\theta^{{\rm sing.}}+\theta^{{\rm reg.}}$, where $\theta^{{\rm sing.}}(x)$ 
obeys the equation (see e.g. [13])  
 
$$\varepsilon_{\mu\nu\lambda\rho}\partial_\lambda
\partial_\rho\theta^{{\rm sing.}}(x)=2\pi\Sigma_{\mu\nu}(x) \eqno (9)$$ 
and 
describes a given electric string configuration,
whereas $\theta^{\rm reg.}(x)$ 
stands for a single-valued fluctuation around this configuration. 
Notice also, that as it has been shown in Ref. [17], the 
integration measure over the field $\theta$ factorizes into the 
product of measures over the fields $\theta^{\rm sing.}$ and 
$\theta^{\rm reg.}$.

Performing   
the path integral duality transformation of Eq. (8) along 
the lines described in Ref. [13], we get 

$${\cal Z}=\exp\left(-(2\pi e)^2\int d^4x\Sigma_{\mu\nu}^2\right)
\int {\cal D}B_\mu \int {\cal D}h_{\mu\nu}\int {\cal D}x_\mu(\xi) 
\exp\left\{\int d^4x\left[-\frac1{12\eta^2}H_{\mu\nu
\lambda}^2+i\pi h_{\mu\nu}\Sigma_{\mu\nu}-\right.\right.$$

$$\left.\left.-\frac14 F_{\mu\nu}^2-
i\tilde F_{\mu\nu}\left(gh_{\mu\nu}+2\pi ie\Sigma_{\mu\nu}
\right)\right]\right\}, \eqno (10)$$
where 
$H_{\mu\nu\lambda}\equiv\partial_\mu h_{\nu\lambda}+
\partial_\lambda h_{\mu\nu}+\partial_\nu h_{\lambda\mu}$ is the field 
strength 
tensor of an antisymmetric tensor field $h_{\mu\nu}$ (the so-called 
Kalb-Ramond field). 
Next, by carrying out the integration over the field $B_\mu$ in 
Eq. (10), we 
obtain

$${\cal Z}=\int {\cal D}x_\mu(\xi)
\int {\cal D}
h_{\mu\nu}\exp \left\{\int d^4x\left[-\frac{1}{12\eta^2}H_{\mu\nu
\lambda}^2
-\frac{1}{4e^2}h_{\mu\nu}^2-i\pi h_{\mu\nu}\Sigma_{\mu\nu}
\right]\right\}.\eqno (11)$$
The details of the 
derivation of Eqs. (10) and (11) are outlined in the Appendix 1.

Finally, the Gaussian 
integration over the field $h_{\mu\nu}$ in Eq. (11) (see 
Appendix 2) 
leads to the following expression for the  
partition function (8)  

$${\cal Z}=
\int {\cal D}x_\mu(\xi)\exp\left\{-\pi^2\int\limits_\Sigma^{} 
d\sigma_{\lambda\nu}(x)
\int\limits_\Sigma^{} d\sigma_{\mu\rho}(y)
D_{\lambda\nu, 
\mu\rho}(x-y)
\right\}. \eqno (12)$$
In Eq. (12), the propagator of the field $h_{\mu\nu}$ 
has the following form

$$D_{\lambda\nu, \mu\rho}(x)\equiv D_{\lambda\nu, \mu\rho}^{(1)}(x)+
D_{\lambda\nu, \mu\rho}^{(2)}(x),$$
where

$$D_{\lambda\nu, \mu\rho}^{(1)}(x)=\frac{\eta^3}{8\pi^2e}\frac
{K_1}{\left|x\right|}\Biggl(\delta_{\lambda\mu}\delta_{\nu\rho}-
\delta_{\mu\nu}\delta_{\lambda\rho}\Biggr), \eqno (13)$$

$$D_{\lambda\nu, \mu\rho}^{(2)}(x)=\frac{e\eta}{4\pi^2x^2}\left\{\Biggl[
\frac{K_1}{\left|x\right|}+\frac m2\left(K_0+K_2\right)\Biggr]
\Biggl(\delta_{\lambda\mu}\delta_{\nu\rho}-\delta_{\mu\nu}\delta_{\lambda
\rho}\Biggr)+\right.$$

$$+\frac{1}{2\left|x\right|}\Biggl[3\Biggl(\frac{m^2}{4}
+\frac{1}{x^2}\Biggr)K_1+\frac{3m}{2\left|x\right|}\left(K_0+
K_2\right)+\frac{m^2}{4}K_3\Biggr]\cdot$$

$$\left.\cdot\Biggl(\delta_{\lambda\rho}x_\mu x_\nu+\delta_{\mu\nu}
x_\lambda 
x_\rho-\delta_{\mu\lambda}x_\nu x_\rho-\delta_{\nu\rho}x_\mu x_\lambda
\Biggr)\right\}. \eqno (14)$$
Here $K_i\equiv K_i(m\left|x\right|),{\,}i=0,1,2,3,$ 
stand for the modified 
Bessel functions, and $m\equiv\frac{\eta}{e}$ 
is the mass of the dual gauge boson 
generated by the Higgs mechanism. 
Due to the Stokes theorem, the term 

$$\int\limits_\Sigma^{} 
d\sigma_{\lambda\nu}(x)\int\limits_\Sigma^{} 
d\sigma_{\mu\rho}(y)D_{\lambda\nu, 
\mu\rho}^{(2)}(x-y)$$
can be rewritten as a boundary one (see Appendix 2), 
which finally leads to the 
following representation for the partition function of extended DAHM 
in the London limit \footnote{ 
It is worth mentioning, that an analogous expression for the partition 
function has been obtained and investigated in Ref. [11]. 
This has been done by making use of techniques  
different from the ones applied in the present paper.} 

$${\cal Z}=\exp\left[-\frac{e\eta}{2}\oint\limits_C^{}dx_\mu
\oint\limits_C^{}dy_\mu\frac{K_1(m|x-y|)}{|x-y|}\right]\cdot$$

$$\cdot
\int{\cal D}x_\mu(\xi)\exp\left[-\frac{\eta^3}{4e}\int\limits_\Sigma^{}
d\sigma_{\mu\nu}(x)\int\limits_\Sigma^{}d\sigma_{\mu\nu}(y)
\frac{K_1(m|x-y|)}{|x-y|}\right]. \eqno (15)$$

The first exponent on the R.H.S. of Eq. (15) leads to the 
short-range Yukawa potential, 

$$V_{\rm Yuk.}(R)\propto\frac{1}{R}{\rm e}^{-mR}.$$
Notice that since quarks and antiquarks were from the very beginning 
considered as classical particles, the external contour $C$  
explicitly enters the final result. 
Would one consider them on the quantum level, Eq. (15) 
must be supplied by a certain prescription of the summation over the 
contours [1,11]. 

The integral over string world-sheets on the R.H.S. of 
Eq. (15) is the essence of the string representation of the 
partition function. Being carried out in the saddle-point 
approximation, it results in the last exponent on the R.H.S. of 
Eq. (15), where the integrals are taken over the surface of the 
minimal area, encircled by the contour $C$, $\Sigma_{\rm min.}$ 
(cf. Eq. (4)). This term then yields a 
rising confining quark-antiquark potential (2).  

As it has already been discussed in the Introduction, the string 
tension $\sigma$ in Eq. (2) is nothing else,  
but the coupling constant of the Nambu-Goto term, 

$$S_{\rm NG}=\sigma\int d^2\xi\sqrt{\hat g},$$
where $\hat g$ stands for the determinant of the induced metric tensor 
$\hat g_{ab}(\xi)=(\partial_ax_\mu(\xi))(\partial_bx_\mu(\xi))$ 
of the string 
world-sheet, $a,b=1,2$. This term is 
the first {\it local} term in the derivative expansion 
of the full {\it nonlocal} string effective action 

$$S_{\rm eff.}=\frac{\eta^3}{4e}\int\limits_{\Sigma_{\rm min.}}^{}
d\sigma_{\mu\nu}(x)\int\limits_{\Sigma_{\rm min.}}^{}d\sigma_{\mu\nu}(y)
\frac{K_1(m|x-y|)}{|x-y|}.$$
The second local term in this expansion is 
the so-called rigidity term [19,20] 

$$S_{\rm rigidity}=
\frac{1}{\alpha_0}
\int d^2\xi\sqrt{\hat g}\hat g^{ab}(\partial_a t_{\mu\nu}(\xi))
(\partial_b t_{\mu\nu}(\xi)),$$
where $t_{\mu\nu}(\xi)=\frac{1}{\sqrt{\hat g}}\varepsilon^{ab} 
(\partial_ax_\mu(\xi))
(\partial_bx_\nu(\xi))$ is the extrinsic curvature 
tensor of the string world-sheet, and $\frac{1}{\alpha_0}$ is the inverse 
bare coupling constant.

Making use of the results of Ref. [5], it 
is possible to derive from the nonlocal 
string effective action $S_{\rm eff.}$ both 
the string tension of the Nambu-Goto term and the inverse bare 
coupling constant of the rigidity term. The latter one occurs 
to be finite and reads 

$$\frac{1}{\alpha_0}=-\frac{\pi e^2}{8}. \eqno (16)$$
In particular, one can see that $\frac{1}{\alpha_0}<0$, which reflects 
the stability of strings [20,15,21] \footnote{Notice, 
that the negative sign 
of this coupling constant alone does not guarantee stability of strings. 
In the non-Abelian case, one should elaborate out another mechanisms 
to ensure this stability and get rid of crumpling of the string 
world-sheet [19,7].}. 
As far as the string tension is concerned, one obtains  

$$\sigma=\pi\eta^2 K_0(a)\simeq
\pi\eta^2\ln\frac{2}{\gamma a}, \eqno (17)$$
where $\gamma=1.781...$ is the Euler's constant, and  $a$ 
stands for a characteristic small dimensionless parameter. 
In the London limit, we get 
$a\sim\frac{m}{M}$, where $M=2\sqrt{2\lambda}\eta$ is the 
monopole mass following from Eq. (6). This mass plays the role of the 
UV momentum cutoff somehow analogous to the inverse lattice 
spacing in QCD. 
Notice, that the logarithmic divergency of the string tension 
in the Ginzburg-Landau model and AHM is a well known result, which 
can be obtained directly from the definition of this quantity as a free 
energy per unit length of the string    
(see e.g. [22]). Clearly, both the string tension 
and the inverse bare coupling constant of the rigidity term are 
not proportional 
to $g$, which means that these quantities are essentially 
nonperturbative similarly to the QCD case (cf. Eq. (3)).

\vspace{6mm}
{\large \bf 3. String Representation for the Generating Functional  
of Field Strength Correlators}
\vspace{3mm}

In this Section, we shall derive the string representation for the 
generating functional of field strength correlators in the 
London limit of extended DAHM. From this we shall then 
obtain an expression 
for the bilocal correlator and compare it with the one in 
QCD. Our starting expression for the generating functional 
reads as follows 

$${\cal Z}\left[S_{\alpha\beta}\right]=
\int {\cal D}B_\mu{\cal D}\theta^{{\rm sing.}}{\cal D}
\theta^{{\rm reg.}}\exp\left\{-\int d^4x\left[\frac14
\left(F_{\mu\nu}-F_{\mu\nu}^E\right)^2+\frac{\eta^2}{2}
\left(\partial_\mu\theta-2gB_\mu\right)^2+iS_{\mu\nu}\tilde 
F_{\mu\nu}\right]\right\}, \eqno (18)$$
where $S_{\mu\nu}$ is a source of the field strength tensor 
$\tilde F_{\mu\nu}=\frac12\varepsilon_{\mu\nu\lambda\rho}
F_{\lambda\rho}$, which obviously corresponds to the field      
strength of the usual gauge field $A_\mu$ in the AHM. 
Performing the same transformations which led from 
Eq. (8) to Eq. (12), we obtain from Eq. (18) 

$${\cal Z}\left[S_{\alpha\beta}\right]=
\exp\left[-\int d^4x\left(S_{\mu\nu}^2+4\pi ieS_{\mu\nu}
\Sigma_{\mu\nu}\right)\right]\cdot$$

$$\cdot\int{\cal D}x_\mu(\xi)\exp\left\{-\int d^4x d^4y
\left(\pi\Sigma_{\lambda\nu}(x)-\frac{i}{e}S_{\lambda\nu}(x)\right)
D_{\lambda\nu, \mu\rho}(x-y)\left(\pi\Sigma_{\mu\rho}(y)-
\frac{i}{e}S_{\mu\rho}(y)\right)\right\}. \eqno (19)$$
Let us now derive from the general form (19) of the 
generating functional 
the expression for the bilocal 
correlator of the field strength tensors. 
The result reads

$$\left.\left<\tilde F_{\lambda\nu}(x)\tilde F_{\mu\rho}(y)\right>=
\frac{1}{{\cal Z}[0]}\frac{\delta^2{\cal Z}\left[S_{\alpha\beta}\right]}
{\delta S_{\lambda\nu}(x)\delta S_{\mu\rho}(y)}
\right|_{S_{\alpha\beta}=0}=$$

$$=\left(\delta_{\lambda\mu}\delta_{\nu\rho}-\delta_{\lambda\rho}
\delta_{\mu\nu}\right)\delta(x-y)+
\frac{2}{e^2}D_{\lambda\nu, \mu\rho}(x-y)-$$

$$-4\pi^2\left<\left(2e\Sigma_{\lambda\nu}(x)-\frac{1}{e}
\int\limits_\Sigma^{}d\sigma_{\alpha\beta}(z)D_{\alpha\beta, 
\lambda\nu}(z-x)\right) 
\left(2e\Sigma_{\mu\rho}(x)-\frac{1}{e}
\int\limits_\Sigma^{}d\sigma_{\gamma\zeta}(u)D_{\gamma\zeta, 
\mu\rho}(u-y)\right)\right>_{x_\mu(\xi)}, \eqno (20)$$
where

$$\left<...\right>_{x_\mu(\xi)}\equiv\frac{\int {\cal D}x_\mu(\xi)(...)
\exp\left[-\pi^2\int\limits_\Sigma^{}d\sigma_{\alpha\beta}(z)
\int\limits_\Sigma^{}d\sigma_{\gamma\zeta}(u)D_{\alpha\beta, \gamma\zeta}
(z-u)\right]}{\int {\cal D}x_\mu(\xi)
\exp\left[-\pi^2\int\limits_\Sigma^{}d\sigma_{\alpha\beta}(z)
\int\limits_\Sigma^{}d\sigma_{\gamma\zeta}(u)D_{\alpha\beta, \gamma\zeta}
(z-u)\right]}$$
is the average over the string world-sheets, 
and the term with the $\delta$-function 
on the R.H.S. of Eq. (20) corresponds to the free contribution to the 
correlator.

Let us next study the large distance asymptotics of Eqs. 
(13) and (14), i.e. 
consider these equations at 
$\left|x\right|\gg\frac{1}{m}$. Then    
one can see, that due to the large distance 
asymptotics 
of the modified Bessel functions, the propagator 
$D_{\lambda\nu, \mu\rho}(x)$ has 
the order of magnitude $\frac{\eta^4}{e^2}$. Therefore, at such distances 
the absolute value of the 
second term on the R.H.S. of Eq. (20) is much larger than the 
absolute value of the last term,  
provided that the following inequality holds 

$$\frac{\eta^2\left|\Sigma\right|}{e}\ll 1, \eqno (21)$$ 
where 
$\left|\Sigma\right|$ stands for the area of the surface 
$\Sigma$. In what follows, we shall restrict ourselves to 
the case of small enough $\eta$ and $g$, for which this inequality 
is valid, and consequently the last term on the R.H.S. of Eq. (20) 
can be disregarded w.r.t. the second one.

Following the SVM [2,8], let us parametrize the bilocal 
correlator of the field strength tensors 
by the two Lorentz structures similarly to Eq. (5) 

$$\left<\tilde F_{\lambda\nu}(x)\tilde F_{\mu\rho}(0)\right>=
\Biggl(\delta_{\lambda\mu}\delta_{\nu\rho}-\delta_{\lambda\rho}
\delta_{\nu\mu}\Biggr)D\left(x^2\right)+$$

$$+\frac12\Biggl[\partial_\lambda
\Biggl(x_\mu\delta_{\nu\rho}-x_\rho\delta_{\nu\mu}\Biggr)
+\partial_\nu\Biggl(x_\rho\delta_{\lambda\mu}-x_\mu\delta_{\lambda\rho}
\Biggr)\Biggr]D_1\left(x^2\right). \eqno (22)$$
Then in the approximation 
(21), by virtue of Eqs. (13) and (14), we arrive at the following 
expressions for the functions $D$ and $D_1$

$$D\left(x^2\right)=\frac{m^3}{4\pi^2}\frac{K_1}{\left|x\right|},  
\eqno (23)$$
and 

$$D_1\left(x^2\right)=\frac{m}{2\pi^2x^2}\Biggl[\frac{K_1}{\left|x\right|}
+\frac{m}{2}\Biggl(K_0+K_2\Biggr)\Biggr]. \eqno (24)$$
In Eq. (23), we have discarded the free $\delta$-function type 
contribution, since we are working at large distances so that 
$x\ne 0$. 
The asymptotic behaviour of the coefficient functions (23) and (24) 
in the limit $\left|x\right|\gg\frac1m$ under study is then given by

$$D\longrightarrow\frac{m^4}{4\sqrt{2}\pi^{\frac32}}
\frac{{\rm e}^{-m\left|x\right|}}{\left(m\left|x\right|\right)^
{\frac32}}, \eqno (25)$$
and 

$$D_1\longrightarrow\frac{m^4}{2\sqrt{2}\pi^{\frac32}}
\frac{{\rm e}^{-m\left|x\right|}}{\left(m\left|x\right|\right)^
{\frac52}}. \eqno (26)$$
For bookkeeping purposes, let us also list here the asymptotic 
behaviour of the functions (23) and (24) 
in the opposite case, $\left|x\right|
\ll\frac1m$. It reads

$$D\longrightarrow\frac{m^2}{4\pi^2x^2}, \eqno (27)$$
and

$$D_1\longrightarrow\frac{1}{\pi^2\left|x\right|^4}. \eqno (28)$$

One can now see that according to the lattice data [9] 
the asymptotic behaviours (25) and (26) are very similar 
to the large distance ones of the nonperturbative parts of the 
functions $D^{\rm QCD}$ and $D_1^{\rm QCD}$, 
which parametrize the 
gauge-invariant bilocal correlator of gluonic field strength tensors in 
QCD (see Eq. (5)). In particular, both 
functions decrease exponentially, and the function $D$ is much larger 
then the function $D_1$ due to the preexponential power-like behaviour.
We also see that the dual gauge boson mass $m$ corresponds to the 
inverse correlation length of the vacuum $T_g^{-1}$. In particular, 
in the string limit of QCD, studied in Refs. [5-7], when $T_g\to 0$ 
while the value of the string tension is kept fixed, $m$ corresponds to 
$\sqrt{\frac{D^{\rm QCD}(0)}{\sigma}}$.

Moreover, the short distance asymptotic behaviours (27) and (28) are also 
in line with the results obtained within the SVM of QCD in 
the lowest order 
of perturbation theory. Namely, at such distances the 
function $D_1^{\rm QCD}$ to the lowest order 
also behaves as $\frac{1}{\left|x\right|^4}$ (which is just due to the 
one-gluon-exchange contribution) 
and is 
much larger than the function $D^{\rm QCD}$ to the same order. 
Let us however 
stress once more, that Eqs. (27) and (28) contain only a part of the 
full information about the asymptotic behaviours of the 
functions $D$ and $D_1$ at small distances. The remaining information 
is contained in the omitted last term on the R.H.S. of Eq. (20), which 
at small distances might become important and modify the 
asymptotic behaviours (27) and (28). 

The above mentioned similarity in the large- and short distance 
asymptotic behaviours of the functions $D$ and $D_1$, which 
parametrize the bilocal correlator of the field strength tensors 
in DAHM and the gauge-invariant correlator in QCD, 
thus supports the original conjecture 
by 't Hooft and Mandelstam concerning the dual Meissner nature of 
confinement.  

\vspace{6mm}
{\large \bf 4. String Representation for the Generating Functional  
of the Monopole Current Correlators}
\vspace{3mm}

In this Section, we shall present the string representation for the 
generating functional of the monopole current correlators in the 
London limit of extended DAHM. Such a 
representation can be derived by virtue of the same path integral 
duality transformation studied above. 
After that, we shall get from the obtained 
generating functional the correlator of two monopole currents and by 
making use of it 
rederive via the equations of motion the coefficient 
function $D$ in the 
bilocal correlator of the field strength tensors. 

In the 
London limit, the generating functional of the monopole currents reads

$$\hat {\cal Z}\left[J_\mu\right]=
\int {\cal D}B_\mu {\cal D}\theta^{{\rm sing.}}{\cal D}\theta^{{\rm reg.}}
\exp
\left\{\int d^4x\left[-\frac14 \left(F_{\mu\nu}-
F_{\mu\nu}^E\right)^2-\frac{\eta^2}{2}\left(
\partial_\mu\theta-2gB_\mu\right)^2+J_\mu j_\mu\right]\right\},$$
where $j_\mu\equiv -2g\eta^2(\partial_\mu\theta-2gB_\mu)$ is just the 
magnetic monopole current \footnote{Rigorously speaking, this is a current 
of the monopole Cooper pairs.}. 

Performing the duality transformation, we get the following 
string representation for $\hat {\cal Z}
\left[J_\mu\right]$

$$\hat {\cal Z}\left[J_\mu\right]=\exp\left[
\frac{m^2}{2}\int d^4x J_\mu^2(x)\right]
\int {\cal D}x_\mu (\xi)
\exp\left[-\pi^2\int\limits_\Sigma^{}d\sigma_{\alpha\beta}(z)
\int\limits_\Sigma^{}d\sigma_{\gamma\zeta}(u)D_{\alpha\beta, \gamma\zeta}
(z-u)\right]\cdot$$

$$\cdot\exp\Biggl\{2g\varepsilon_{\lambda\nu\alpha\beta}
\int d^4x d^4y\Biggl[
-\frac{g}{2}\varepsilon_{\mu\rho\gamma\delta}\Biggl(\frac{\partial^2}
{\partial x_\alpha\partial y_\gamma}D_{\lambda\nu, \mu\rho}(x-y)
\Biggr)J_\beta (x) J_\delta (y)+$$

$$+\pi \Sigma_{\mu\rho}(y)\Biggl(
\frac{\partial}{\partial x_\alpha} D_{\lambda\nu, \mu\rho}(x-y)\Biggr)
J_\beta (x)\Biggr]\Biggr\}. \eqno (29)$$
Varying now Eq. (29) 
twice w.r.t. $J_\mu$, setting then $J_\mu$ equal to zero, 
and dividing the result by $\hat {\cal Z}[0]$, we arrive 
at the following expression for the correlator of two monopole currents  

$$\left<j_\beta(x) j_\sigma(y)\right>=m^2\delta_{\beta\sigma}\delta(x-y)+
4g^2
\varepsilon_{\lambda\nu\alpha\beta}\varepsilon_
{\mu\rho\gamma\sigma}\Biggl[-\frac12\frac{\partial^2}{\partial 
x_\alpha \partial y_\gamma}D_{\lambda\nu, \mu\rho}(x-y)+$$

$$+\pi^2\left< 
\int\limits_\Sigma^{} d\sigma_{\delta\zeta}(z)\int\limits_\Sigma^{} 
d\sigma_{\chi\varphi}(u)
\Biggl(\frac{\partial}{\partial x_\alpha}D_{\lambda\nu, 
\delta\zeta}(x-z)\Biggr)\Biggl(\frac{\partial}
{\partial y_\gamma} D_{\mu\rho, \chi\varphi}(y-u)\Biggr)
\right>_{x_\mu(\xi)}\Biggr]. \eqno (30)$$

It is straightforward to see that the contribution of the 
term (14) to the R.H.S. of Eq. (30) vanishes, whereas the contribution of 
Eq. (13) to the second term in the square brackets on the R.H.S. 
of Eq. (30) can be disregarded w.r.t. 
its contribution to the first term, provided that the inequality (21) holds. 
Within this 
approximation, making use of the 
equation [2]

$$\left<j_\beta(x)j_\sigma(y)\right>=\Biggl(\frac{\partial^2}
{\partial x_\mu\partial y_\mu}\delta_{\beta\sigma}-
\frac{\partial^2}{\partial x_\beta\partial y_\sigma}\Biggr)
D\left((x-y)^2\right), \eqno (31)$$
which follows from Eq. (22) due to equations of motion, 
we recover from Eqs. (30), (31), and (13) the expression 
for the function $D$ given by Eq. (23).

Notice in conclusion, that only 
the function $D$ can be obtained from the correlator (30) 
due to the 
independence of the latter of the function $D_1$.  

\vspace{6mm}
{\large \bf 5. Summary and Outlook}
\vspace{3mm}

In the present paper, we have derived the string representation 
for the partition function of DAHM extended by introducing gauge fields 
of external currents of electrically charged particles (``quarks''),   
in the London limit. Such a representation yielded the confining  
and the Yukawa parts of the quark-antiquark interaction potential. 
By the local expansion of the obtained nonlocal string effective action, 
one gets, in particular, the corresponding expressions for 
the string tension of the 
Nambu-Goto term and the inverse bare coupling constant of the rigidity 
term. Those occured to be positive and negative respectively, which 
confirms the stability of strings (cf. also [20,15,21]). 
Besides that, both of these 
quantities are not proportional to the coupling constant, 
which means that the string nature of DAHM is of the same 
nonperturbative kind as the one of QCD.

In Section 3, we have obtained the string representation for the 
generating functional of the field strength 
correlators. In a certain approximation (see Eq. (21)), this 
generating functional determined then the expressions 
of the two coefficient functions,  
$D$ and $D_1$ (cf. Eqs. (23) and (24)), 
which parametrize the bilocal correlator. Those occured to 
be quite similar to the corresponding functions in QCD. 
In particular, it turned out that 
the large distance 
asymptotic behaviour of the obtained functions in the extended DAHM was   
in agreement with the existing lattice data [9] 
concerning the corresponding 
behaviours of the nonperturbative parts of these functions in QCD. 
We have also argued that the mass of the dual gauge boson in our 
approach corresponds to the inverse correlation length of the vacuum 
in QCD.  
These results together with SVM 
support the 't Hooft-Mandelstam conjecture 
about the dual Meissner nature of confinement. 

Finally, in Section 4, 
we have obtained the string representation for the generating functional 
of monopole current correlators in the extended DAHM. Then,  
by making use of the equations of motion, 
and the string representation for the correlator of two monopole 
currents,  
we have rederived the coefficient function $D$, which confirms the 
correctness of both 
approaches. 

Thus, we have demonstrated the relevance of the extended DAHM for 
the description of the string properties of confinement in QCD 
according to SVM and the 
't Hooft-Mandelstam scenario. And vice versa, 
our results also 
support the validity of the bilocal approximation in SVM of QCD 
and provide us with some new insights concerning the structure of 
the QCD vacuum.   
In conclusion, all this shows the usefulness of the 
string representation of extended DAHM obtained on the basis of 
the path integral duality transformation. 

Clearly, it is now a challenge to apply this approach to the 
realistic non-Abelian case of QCD. In the spirit of this paper, 
it is in particular interesting to study dual QCD in the framework 
of the so-called Abelian projection method [23,18]. Work in this 
direction is now in progress.

\vspace{6mm}
{\large \bf 6. Acknowledgments}
\vspace{3mm}

We are indebted to A. Di Giacomo, H. Dorn, H. Kleinert, 
M.I. Polikarpov, Chr. Preitschopf, and Yu.A. Simonov for 
useful discussions. 
One of us (D.A.) would also like to thank   
the theory group of the 
Quantum Field Theory Department of the Institute of Physics of the 
Humboldt University of Berlin for kind hospitality and Graduiertenkolleg 
{\it Elementarteilchenphysik} for financial support. 

\vspace{6mm}
{\large \bf Appendix 1. Derivation of Eqs. (10) and (11).}
\vspace{3mm}

In this Appendix, we shall outline some details of the derivation of 
Eqs. (10) and (11). 
Firstly, one can linearize the term $\frac{\eta^2}{2}
\left(\partial_\mu\theta-2gB_\mu
\right)^2$ in the exponent on the R.H.S. of Eq. (8) 
and carry out the integral over 
$\theta^{{\rm reg.}}$ as follows 

$$\int {\cal D}\theta^{{\rm reg.}}\exp\left\{-\frac{\eta^2}{2}
\int d^4x \left(\partial_\mu\theta-2gB_\mu
\right)^2\right\}=$$

$$=\int {\cal D}C_\mu {\cal D}\theta^{{\rm reg.}}
\exp\left\{\int d^4x\left[-\frac{1}{2\eta^2}C_\mu^2+iC_\mu
\left(\partial_\mu\theta-2gB_\mu
\right)\right]\right\}=$$

$$=\int {\cal D}C_\mu\delta\left(\partial_\mu C_\mu\right)
\exp\left\{\int d^4x\left[-\frac{1}{2\eta^2}C_\mu^2+iC_\mu
\left(\partial_\mu\theta^{{\rm sing.}}-2gB_\mu
\right)\right]\right\}. \eqno (A1.1)$$
The constraint $\partial_\mu C_\mu=0$ can be uniquely resolved by 
representing $C_\mu$ in the form $C_\mu=\frac12
\varepsilon_{\mu\nu\lambda\rho}\partial_\nu h_{\lambda\rho}$, where 
$h_{\lambda\rho}$ stands for an antisymmetric tensor field. 
Notice, that the number 
of degrees of freedom during such a replacement 
is conserved, since both of the fields 
$C_\mu$ and $h_{\mu\nu}$ have three independent components.

Then, taking 
into account the relation (9) between $\theta^{{\rm sing.}}$ and 
$\Sigma_{\mu\nu}$, we get 
from Eq. (A1.1) 

$$\int {\cal D}\theta^{{\rm sing.}}
{\cal D}\theta^{{\rm reg.}}\exp\left\{-\frac{\eta^2}{2}
\int d^4x \left(\partial_\mu\theta-2gB_\mu
\right)^2\right\}=$$

$$=\int {\cal D}h_{\mu\nu} {\cal D}x_\mu (\xi)\exp\left\{\int d^4x 
\left[-\frac{1}{12\eta^2}H_{\mu\nu\lambda}^2+i\pi h_{\mu\nu}
\Sigma_{\mu\nu}-ig\varepsilon_{\mu\nu\lambda\rho}B_\mu
\partial_\nu h_{\lambda\rho}\right]\right\}. \eqno (A1.2)$$
In the derivation of Eq. (A1.2), 
we have replaced ${\cal D}\theta^{{\rm sing.}}$ by 
${\cal D}x_\mu(\xi)$ (since the surface $\Sigma$, parametrized by 
$x_\mu(\xi)$, is just the surface, at which the field 
$\theta$ is singular) and, for 
simplicity, have discarded the Jacobian arising during 
such a change of the integration variable \footnote{For the case when 
the surface $\Sigma$ has a spherical topology, this Jacobian 
has been calculated in Ref. [17].}. 

Bringing now together Eqs. (8) and (A1.2), we arrive at Eq. (10). 
In the literature, the above described sequence of 
transformations 
of integration variables is   
usually called ``path integral duality transformation''. In particular, it 
has been 
applied in Ref. [13] to the model with a {\it global} $U(1)$-symmetry.   

Let us now derive Eq. (11). 
To this end, we find it convenient 
to rewrite  

$$\exp\left(-\frac14\int d^4x 
F_{\mu\nu}^2\right)=
\int {\cal D} G_{\mu\nu}\exp\left\{\int d^4x
\left[-G_{\mu\nu}^2+i\tilde F_{\mu\nu}
G_{\mu\nu}\right]\right\},$$
after which the $B_\mu$-integration in Eq. (10) yields 

$$\int {\cal D}B_\mu \exp\left\{-\int d^4x\left[\frac14 F_{\mu\nu}^2+
i\tilde F_{\mu\nu}\left(gh_{\mu\nu}+2\pi ie\Sigma_{\mu\nu}
\right)\right]\right\}=$$

$$=\int {\cal D}G_{\mu\nu}\exp\left(-\int d^4x G_{\mu\nu}^2
\right)\delta\left(\varepsilon_{\mu\nu\lambda\rho}\partial_\mu
\left(G_{\lambda\rho}-gh_{\lambda\rho}-2\pi ie\Sigma_{\lambda\rho}\right)
\right)=$$

$$=\int {\cal D}\Lambda_\mu\exp\left[-\int d^4x\left(gh_{\mu\nu}+2\pi ie
\Sigma_{\mu\nu}
+\partial_\mu \Lambda_\nu-\partial_\nu \Lambda_\mu\right)^2\right]. 
\eqno (A1.3)$$
In the last line of Eq. (A1.3), the constraint 

$$\varepsilon_{\mu\nu\lambda\rho}\partial_\mu
\left(G_{\lambda\rho}-gh_{\lambda\rho}-2\pi ie\Sigma_{\lambda\rho}\right)
=0$$
has been resolved by setting $G_{\lambda\rho}=gh_{\lambda\rho}+
2\pi ie\Sigma_{\lambda\rho}+\partial_\lambda \Lambda_\rho-
\partial_\rho \Lambda_\lambda$.

Finally, by performing in Eq. (A1.3) 
the hypergauge transformation $h_{\mu\nu}\to h_{\mu\nu}+
\partial_\mu\lambda_\nu-\partial_\nu\lambda_\mu$ 
and fixing the gauge by choosing 
$\lambda_\mu=-\frac{1}
{g}\Lambda_\mu$ (see e.g. Ref. [15]), we arrive, omitting the measure 
factor ${\cal D}\Lambda_\mu$, at Eq. (11). 

\vspace{6mm}
{\large \bf Appendix 2. Integration over the Kalb-Ramond Field in 
Eq. (11).}
\vspace{3mm}

Let us carry out the following integration over the Kalb-Ramond field  

$${\cal Z}=\int Dh_{\mu\nu}\exp\Biggl[
\int dx\Biggl(-\frac{1}
{12\eta^2}
H_{\mu\nu\lambda}^2-\frac{1}{4e^2}h_{\mu\nu}^2-
i\pi h_{\mu\nu}\Sigma_{\mu\nu}
\Biggr)\Biggr]. \eqno (A2.1)$$ 
To this end, it is necessary to substitute the saddle-point value of the 
integral (A2.1) 
back into the integrand. The saddle-point equation in the momentum 
representation reads 

$$\frac{1}{2\eta^2}\left(p^2h_{\nu\lambda}^{\rm extr.}(p)+p_\lambda 
p_\mu h_{\mu\nu}^{\rm extr.}(p)+p_\mu p_\nu h_{\lambda\mu}^{\rm extr.}(p)
\right)+\frac{1}{2e^2}h_{\nu\lambda}^{\rm extr.}(p)=-i\pi
\Sigma_{\nu\lambda}(p).$$
This equation can be most easily solved by rewriting it in the 
following way

$$\left(p^2 {\bf P}_{\lambda\nu, \alpha\beta}+m^2 {\bf 1}_{\lambda\nu, 
\alpha\beta}\right)h_{\alpha\beta}^{\rm extr.}(p)=-2\pi i\eta^2
\Sigma_{\lambda
\nu}(p), \eqno (A2.2)$$
where we have introduced the following projection operators 

$${\bf P}_{\mu\nu, \lambda\rho}\equiv\frac12\left({\cal P}_{\mu\lambda}
{\cal P}_{\nu\rho}-{\cal P}_{\mu\rho} {\cal P}_{\nu\lambda}\right)$$
and

$${\bf 1}_{\mu\nu, \lambda\rho}\equiv\frac12 \left(\delta_{\mu\lambda}
\delta_{\nu\rho}-\delta_{\mu\rho}\delta_{\nu\lambda}\right)$$
with ${\cal P}_{\mu\nu}\equiv\delta_{\mu\nu}-\frac{p_\mu p_\nu}{p^2}$. 
These projection operators obey the following relations 

$${\bf 1}_{\mu\nu, \lambda\rho}=-{\bf 1}_{\nu\mu, \lambda\rho}=
-{\bf 1}_{\mu\nu, \rho\lambda}={\bf 1}_{\lambda\rho, \mu\nu}, 
\eqno (A2.3)$$

$${\bf 1}_{\mu\nu, \lambda\rho} {\bf 1}_{\lambda\rho, \alpha\beta}=
{\bf 1}_{\mu\nu, \alpha\beta} \eqno (A2.4)$$
(the same relations hold for ${\bf P}_{\mu\nu, \lambda\rho}$), and 

$${\bf P}_{\mu\nu, \lambda\rho}\left({\bf 1}-{\bf P}\right)_{\lambda
\rho, \alpha\beta}=0. \eqno (A2.5)$$
By virtue of properties (A2.3)-(A2.5), the solution of Eq. (A2.2) reads 

$$h_{\lambda\nu}^{\rm extr.}(p)=-\frac{2\pi i\eta^2}{p^2+m^2}\left[
{\bf 1}+\frac{p^2}{m^2}\left({\bf 1}-{\bf P}\right)\right]_{\lambda
\nu, \alpha\beta}\Sigma_{\alpha\beta}(p),$$
which, once being substituted back into partition function (A2.1), yields 
for it the following expression

$${\cal Z}=\exp\Biggl\{-\pi^2\eta^2\int\frac{d^4p}{(2\pi)^4}
\frac{1}{p^2+m^2}\left[{\bf 1}+\frac{p^2}{m^2}\left({\bf 1}-{\bf P}
\right)\right]_{\mu\nu, \alpha\beta}\Sigma_{\mu\nu}(-p)
\Sigma_{\alpha\beta}(p)
\Biggr\}. 
\eqno (A2.6)$$
Rewriting Eq. (A2.6) in the coordinate representation we arrive at 
Eq. (12).

Let us now prove that the term 
proportional to the projection operator ${\bf 1}-{\bf P}$ on the 
R.H.S. of Eq. (A2.6) indeed yields in the 
coordinate representation the 
boundary term, i.e. Eq. (14). One has 

$$p^2({\bf 1}-{\bf P})_{\lambda\nu, \alpha\beta}=\frac12
(\delta_{\nu\beta}p_\lambda p_\alpha+\delta_{\lambda\alpha}
p_\nu p_\beta-\delta_{\nu\alpha}p_\lambda p_\beta-\delta_{\lambda
\beta}p_\nu p_\alpha). \eqno (A2.7)$$
By making use of Eq. (A2.7), the term 

$$-\pi^2\eta^2\int\frac{d^4p}{(2\pi)^4}\frac{1}{p^2+m^2}
\frac{p^2}{m^2}({\bf 1}-{\bf P})_{\mu\nu, \alpha\beta}\int d^4x\int d^4y
{\rm e}^{ip(y-x)}\Sigma_{\mu\nu}(x)\Sigma_{\alpha\beta}(y)$$
under study, after carrying out the integration over $p$, reads

$$\frac{\eta^2}{2m}\int d^4x \Sigma_{\mu\nu}(x)
\int d^4y \Sigma_{\nu\beta}(y) 
\frac{\partial^2}{\partial x_\mu\partial y_\beta}\frac{K_1(m\left|x-y
\right|)}{\left|x-y\right|}. \eqno (A2.8)$$
Acting in Eq. (A2.8) straightforwardly with the derivatives, we 
arrive at Eq. (14). 
However, one can perform the partial integration, which gives the 
boundary term  

$$-\frac{e^2m}{2}\oint\limits_C^{}dx_\mu
\oint\limits_C^{}dy_\mu\frac{K_1(m\left|x-y\right|)}{\left|
x-y\right|}, $$
in which one can recognize the argument of the first exponent 
standing on the R.H.S. of Eq. (15).

\newpage
{\large \bf References}

\vspace{3mm}
\noindent
$[1]$~A.M. Polyakov, Gauge Fields and Strings (Harwood Academic 
Publishers, Chur, Switzerland, 1987).\\
$[2]$~Yu.A. Simonov, Phys. Usp. 
{\bf 39} (1996) 313.\\
$[3]$~K.G. Wilson, Phys. Rev. {\bf D 10} (1974) 2445.\\ 
$[4]$~I.-J. Ford et al., Phys. Lett. {\bf B 208} (1988) 286; 
E. Laermann et al., Nucl. Phys. {\bf B} (Proc. Suppl.) {\bf 26} 
(1992) 268.\\
$[5]$~D.V. Antonov, D. Ebert, and Yu.A. Simonov, Mod. Phys. Lett. 
{\bf A 11} (1996) 1905.\\
$[6]$~D.V. Antonov and D. Ebert, Mod. Phys. Lett. {\bf A 12} (1997) 
2047.\\
$[7]$~D.V. Antonov and D. Ebert, preprints HUB-EP-98-15 and 
hep-ph/9802353 (1998) (Phys. Rev. {\bf D}, in press).\\
$[8]$~H.G. Dosch, Phys. Lett. {\bf B 190} (1987) 177; H.G. Dosch and 
Yu.A. Simonov, Phys. Lett. {\bf B 205} (1988) 339; Yu.A. Simonov, 
Nucl. Phys. {\bf B 324} (1989) 67; Sov. J. Nucl. Phys. {\bf 54} 
(1991) 115; H.G. Dosch, Prog. Part. Nucl. Phys. 
{\bf 33} (1994) 121.\\
$[9]$~A. Di Giacomo and H. Panagopoulos, Phys. Lett. {\bf B 285}   
(1992) 133; M. D'Elia, A. Di Giacomo, and E. Meggiolaro, Phys. Lett. 
{\bf B 408} (1997) 315; 
A. Di Giacomo, 
E. Meggiolaro, and H. Panagopoulos, Nucl. Phys. 
{\bf B 483} (1997) 371; Nucl. Phys. Proc. Suppl. {\bf A 54} (1997) 343.\\
$[10]$S. Mandelstam, Phys. Rep. {\bf C 23} (1976) 245;  
G. 't Hooft, in: High Energy Physics, ed. A. Zichichi  
(Editrice Compositori, Bologna, 1976).\\
$[11]$H. Kleinert, Gauge Fields in Condensed Matter, 
Vol. 1: Superflow and Vortex Lines. Disorder Fields, Phase Transitions 
(World Scientific, Singapore, 1989); Phys. Lett. {\bf B 246} (1990) 127; 
ibid. {\bf B 293} 
(1992) 168; 
M. Kiometzis, H. Kleinert, and A.M.J. Schakel, 
Fortschr. Phys. {\bf 43} (1995) 697; H. Kleinert, 
preprint cond-mat/9503030 (1995).\\
$[12]$A.A. Abrikosov, Sov. Phys. JETP {\bf 32} (1957) 1442; 
H.B. Nielsen and P. Olesen, Nucl. Phys. {\bf B 61} (1973) 45.\\
$[13]$K. Lee, Phys. Rev. {\bf D 48} (1993) 2493.\\
$[14]$M.I. Polikarpov, 
U.-J. Wiese, and M.A. Zubkov, Phys. Lett. {\bf B 309}  
(1993) 133.\\ 
$[15]$P. Orland, Nucl. Phys. {\bf B 428} (1994) 221.\\
$[16]$M. Sato and S. Yahikozawa, Nucl. Phys. {\bf B 436} (1995) 100.\\
$[17]$E.T. Akhmedov, M.N. Chernodub, M.I. Polikarpov, and M.A. Zubkov, 
Phys. Rev. 
{\bf D 53} (1996) 2087.\\ 
$[18]$M.N. Chernodub and M.I. Polikarpov, preprints hep-th/9710205 
and ITEP-TH-55-97 (1997).\\ 
$[19]$A.M. Polyakov, Nucl. Phys. {\bf B 268} (1986) 406.\\
$[20]$H. Kleinert, Phys. Lett. {\bf B 174} (1986) 335; ibid. 
{\bf B 211} (1988) 151.\\
$[21]$M.C. Diamantini, F. Quevedo, and C.A. Trugenberger, Phys. Lett. 
{\bf B 396} (1997) 115;  
M.C. Diamantini and C.A. Trugenberger, Phys. Lett. {\bf B 421} 
(1998) 196; preprint hep-th/9803046 (1998).\\
$[22]$E.M. Lifshitz and L.P. Pitaevski, Statistical Physics, Vol. 2 
(Pergamon, New York, 1987).\\
$[23]$G. 't Hooft, Nucl. Phys. {\bf B 190} (1981) 455; 
T. Suzuki et al., Prog. Theor. Phys. {\bf 80} (1988) 929; 
Phys. Lett. {\bf B 294} (1992) 100; 
H. Suganuma et al., Nucl. Phys. {\bf B 435} (1995) 207; see also 
M. Baker, J.S. Ball, and F. Zachariasen, Phys. Rep. {\bf 209} (1991) 73.\\

\end{document}